\documentclass[english,twocolumn,showpacs,prb]{revtex4}
\usepackage[T1]{fontenc}
\usepackage[latin9]{inputenc}
\usepackage{graphicx}
\usepackage{color}
\usepackage{cases}
\begin{document}
\makeatletter
\@ifundefined{textcolor}{}
{%
 \definecolor{BLACK}{gray}{0}
 \definecolor{WHITE}{gray}{1}
 \definecolor{RED}{rgb}{1,0,0}
 \definecolor{GREEN}{rgb}{0,1,0}
 \definecolor{BLUE}{rgb}{0,0,1}
 \definecolor{CYAN}{cmyk}{1,0,0,0}
 \definecolor{MAGENTA}{cmyk}{0,1,0,0}
 \definecolor{YELLOW}{cmyk}{0,0,1,0}
 }
\title{Quantum interference in topological insulator Josephson junctions}
\author{Juntao Song$^{1}$}
\author{Haiwen Liu$^{2,3,4}$}
\author{Jie Liu$^{5}$}
\author{Yu-Xian Li$^{1}$}
\author{Robert Joynt$^{2,6}$}
\author{Qing-feng Sun$^{2,3}$}
\author{X. C. Xie$^{2,3}$}
\affiliation{$^{1}$Department of Physics and Hebei Advanced Thin Film Laboratory, Hebei
Normal University, Shijiazhuang 050024, China}
\affiliation{$^{2}$International Center for Quantum Materials, School of Physics, Peking
University, Beijing 100871, China }
\affiliation{$^{3}$Collaborative Innovation Center of Quantum Matter, Beijing 100871, China }
\affiliation{$^{4}$Department of Physics, Center for Advanced Quantum Studies, Beijing Normal University, Beijing 100875, China}
\affiliation{$^{5}$Department of Applied Physics, School of Science, Xian Jiaotong
University, Xian 710049, China }
\affiliation{$^{6}$Physics Department, University of Wisconsin-Madison, 1150 University
Avenue, Madison, Wisconsin 53706, USA}

\date{\today}

\begin{abstract}
Using non-equilibrium Green's functions, we studied numerically the transport properties of a Josephson junction, superconductor-topological insulator-superconductor hybrid system. Our numerical calculation shows first that proximity-induced superconductivity is indeed observed in the edge states of a topological insulator adjoining two superconducting leads and second that the special characteristics of topological insulators endow the edge states
with an enhanced proximity effect with a superconductor but do not forbid the bulk states to do the same. In a size-dependent analysis of the local current, it was found that a few residual bulk states can lead to measurable
resistance, whereas because these bulk states spread over the whole sample, their contribution to the interference pattern is insignificant when the sample size is in the micrometer range. Based on these numerical results, it is concluded that the apparent disappearance of residual bulk states in the superconducting interference process as described in Ref. [\onlinecite{HartNautrePhys2014f}] is just due to the effects of size: the contribution of the topological edge states outweighs that of the residual bulk states.
\end{abstract}

\maketitle
\section{Introduction}
Since the original theoretical proposals\cite{HaldanePRL1988j,KanePRL2005d,BernevigScience2006j,CXLiuPRL2008h} of the existence of topological insulators (TIs) and their eventual realization in HgTe/CdTe and InAs/GaSb quantum wells (QWs),\cite{KonigScience2007r,RothScience2009w,KnezPRL2014d} extensive studies of their properties have been performed. Because of the topological non-triviality of topological insulators, it was expected that the dissipationless current carried by topologically protected edge states could be utilized to construct electronic devices with low power consumption. Experimentally, however, the edge states that were expected to be robust were found to become unstable when the sample of HgTe/CdTe quantum wells became somewhat larger and reached sizes of a few microns. This experimental finding stirred up a lot of interest. How disorder, magnetic impurity, electron-electron interaction, electron-phonon interaction, etc, and the various combinations of these effects, could affect transport in edge states as well as the topological properties of TIs themselves have been intensively studied.\cite{LiJianPRL2009o,JiangHPRB2009e,BeenakkerPRL2009i,SongJTPRB2012w,YTanakaPRL2011r,HWLiuPRL2014p,JMaciejko2010w,PDelplacePRL2012w}

The notion that topological superconductors (TSs), similar to TIs, possess topologically protected gapless edge or surface states traversing through the superconductor gap, was also put forward. It is anticipated\cite{FuLiangPRL2008a} that superconducting vortices in two dimensional chiral $p$-wave TSs can bind Majorana fermions (MFs), which persist at zero energy and obey non-Abelian braiding statistics. This implies that adiabatically exchanging two Majorana fermions noncommutatively transforms the system from one ground state to another. Due to the protection of the non-Abelian braiding properties a quantum bit encoded in MFs is immune to external environmental noise. This gives MFs considerable promise in topological quantum computing. Motivated by this idea, a great deal of effort has been expended to examine many materials and structures to find ways to produce MFs.

Because MFs can appear when $s$ wave superconductors are put in close proximity to TIs, great attention has recently been focused on this kind of superconducting hybrid system.\cite{SunPRB2011a,IKnezPRL2012f,AdroguerPRB2010f,NarayanPRB2010f,PGhaemiPRL2016t,JLinderPRL2010k,TanakaPRL2009k,CBenjiaminEPL2016r,AZyuzinArXiv2015d} One phenomenon that also appears in superconducting hybrid systems is Andreev reflection (AR), which is an important transport process found about fifty years ago.\cite{AFAndreevSPJETP1964t}  AR occurs near the interface of a conductor and a superconductor: an electron incident from the metallic side is reflected as a hole while a Cooper pair is created in the superconductor. Recently, it has been reported theoretically that AR in TI-superconductor (TI-S) hybrid junctions should manifest quantized characteristics due to topologically protected helical edge states\cite{SunPRB2011a,IKnezPRL2012f} and thus may be used as a powerful method to probe helical edge modes in TI system.\cite{AdroguerPRB2010f} More intriguingly, it is anticipated that the promising existence of MFs in Josephson junctions with a superconductor-TI-superconductor (S-TI-S) structure will make the Josephson supercurrent oscillate with a period of $4\pi$-twice that of conventional junctions.\cite{BeenakkerPRL2013f,SPLeePRL2014f} \ Theoretical proposals focus on correlated phase measurements to prove the existence of MFs in these hybrid systems. This means that the first crucial thing to do is to study the behavior of the helical edge or surface states in close proximity to a superconductor.

To date, many experimental groups have observed proximity-induced superconductivity in the surface or edge states of
TIs.\cite{MVeldhorstNatureMat2012k,HartNautrePhys2014f,PribiagNatNanotechnology2014h} In particular, proximity-induced superconductivity in the quantum spin Hall edge of HgTe/CdTe QWs has been reported, and clear superconducting quantum interference patterns due to the superconducting helical edge states of TIs were obtained independently by two experimental groups.\cite{HartNautrePhys2014f,PribiagNatNanotechnology2014h} Following the Dynes-Fulton analysis process, the supercurrent as a function of transverse position in a S-TI-S junction was obtained and found to be localized on two edges of the junction. However, one large puzzle remained in the discussion in Ref. [\onlinecite{HartNautrePhys2014f}]; namely, that the supercurrent was localized on the two edges of the junction even though the system still had measurable resistance due to residual bulk states, which are supposed to spread widely in whole sample. The Dynes-Fulton approach was discussed in very recent theoretical work,\cite{HuiPRB2014g} where it was pointed out the space-distributed supercurrent obtained through the Frourier transformation of the Dynes-Fulton process is not quantitatively accurate and sometimes deviates significantly from the correct value. Therefore, it is very important to perform a direct numerical simulation on this S-TI-S hybrid junction and clearly verify the existence of proximity-induced superconductivity in the helical edge states of quantum spin Hall systems, and also present reasonable explanations of the experimental results. That is the purpose of the present work.

The rest of this paper is organized as follows: In Sec.~\ref{sec:models}, we introduce the Hamiltonian of superconducting leads and HgTe/CdTe QWs, present the formula for the current through the superconducting leads, as well as for the local current through the cross section of HgTe/CdTe QWs. The numerical results are discussed in Sec.~\ref{sec:discussions}. Finally, a brief summary is given in Sec.~\ref{sec:conclusions}.

\section{Model and Formalism}
\label{sec:models}
\subsection{Hamiltonians for superconducting leads and HgTe/CdTe QWs}\label{sec:Hamiltonians}
\label{sec:Hamiltonians} The two-dimensional TI phase has been experimentally realized in the HgTe/CdTe\cite{KonigScience2007r,RothScience2009w} and InAs/GaSb\cite{KnezPRL2014d} QWs. In order to easily compare our numerical simulation to the experimental results in Refs. [\onlinecite{HartNautrePhys2014f}] and [\onlinecite{PribiagNatNanotechnology2014h}], here we shall adopt the parameters appropriate for HgTe/CdTe QWs in the following calculations. The results should be also qualitatively correct for other 2D TI systems, e.g. InAs/GaSb QWs. We consider a superconductor-HgTe/CdTe QW-superconductor Josephson junction as shown in Fig. {\ref{Figure1}}. This hybrid device is described by the Hamiltonian $H=H_{C}+H_{\scriptscriptstyle{\mathbf{S}}}+H_{T}$, where $H_{C}$, $H_{\scriptscriptstyle{\mathbf{S}}}$, and $H_{T}$ are the Hamiltonians of the HgTe/CdTe QWs, superconducting leads, and the coupling between them, respectively.

In the Nambu representation, the tight binding Hamiltonian $H_{C}$ of the HgTe/CdTe QWs is represented as:
\begin{equation}
H_{C}=\sum\limits_{\mathbf{i}}\Psi_{\mathbf{i}}^{\dagger}H_{\mathbf{i}%
,\mathbf{i}}\Psi_{\mathbf{i}}+\sum\limits_{\mathbf{i},\mathbf{i}%
+\hat{\mathbf{r}}}\Psi_{\mathbf{i}}^{\dagger}f_{\mathbf{i},\mathbf{i}%
+\hat{\mathbf{r}}}H_{\mathbf{i},\mathbf{i}+\hat{\mathbf{r}}}\Psi
_{\mathbf{i}+\hat{\mathbf{r}}}+H.c.,\label{M2}%
\end{equation}
where $\mathbf{i}=(\mathbf{i}_{x},\mathbf{i}_{y})$ is the site index, $\hat{\mathbf{r}}=\hat{\mathbf{x}}$ or $\hat{\mathbf{y}}$ represents the unit vector along $x$ or $y$ direction, $\Psi_{\mathbf{i}}=(c_{s\mathbf{i}\uparrow
},c_{p\mathbf{i}\uparrow},c_{s\mathbf{i}\downarrow}^{\dagger},c_{p\mathbf{i} \downarrow}^{\dagger})^{T}$, and $c_{s\mathbf{i}\uparrow},c_{p\mathbf{i} \uparrow},c_{s\mathbf{i}\downarrow},c_{p\mathbf{i}\downarrow}$ are
annihilation operators of electrons at the site $\mathbf{i}$ in the states $|s,\uparrow\rangle$, $|p_{x}+ip_{y},\uparrow\rangle$, $|s,\downarrow\rangle$, and $|-(p_{x}-ip_{y}),\downarrow\rangle$. In Eq.(\ref{M2}), $H_{\mathbf{i} ,\mathbf{i}/\mathbf{i},\mathbf{i}+\hat{\mathbf{r}}}=\left(
\begin{array}
[c]{ll}
h_{\mathbf{i},\mathbf{i}/\mathbf{i},\mathbf{i}+\hat{\mathbf{r}}} &
\mathbf{0}\\
\mathbf{0} & -h_{\mathbf{i},\mathbf{i}/\mathbf{i},\mathbf{i}+\hat{\mathbf{r}}}
\end{array}
\right)  $ and $f_{\mathbf{i},\mathbf{i}+\hat{\mathbf{r}}}=\left(
\begin{array}
[c]{ll}
e^{i\phi_{\mathbf{i},\mathbf{i}+\hat{\mathbf{r}}}} & \mathbf{0}\\
\mathbf{0} & e^{-i\phi_{\mathbf{i},\mathbf{i}+\hat{\mathbf{r}}}}
\end{array}
\right)  $ are $4\times4$ matrices, where $h_{\mathbf{i},\mathbf{i}}=\left(
\begin{array}
[c]{ll}
E_{s} & 0\\
0 & E_{p}
\end{array}
\right)  $, $h_{\mathbf{i},\mathbf{i}+\hat{\mathbf{x}}}=\left(
\begin{array}
[c]{ll}
t_{ss} & -it_{sp}\\
-it_{sp} & t_{pp}%
\end{array}
\right)  $, and $h_{\mathbf{i},\mathbf{i}+\hat{\mathbf{y}}}=\left(
\begin{array}
[c]{ll}
t_{ss} & -t_{sp}\\
t_{sp} & t_{pp}
\end{array}
\right)  $, with $E_{s/p}=C\pm M-E_{F}-4(D\pm B)/a^{2}$, $t_{ss/pp}=(D\pm B)/a^{2}$ and $t_{sp}=A/2a$. Adopting a definite gauge for the magnetic vector potential ($\mathbf{A}=(yB_{\bot},0,0),\phi_{\mathbf{i},\mathbf{j}}=2\pi
\int_{\mathbf{i}}^{\mathbf{j}}\mathbf{A}{\cdot}d\mathbf{l}/\Phi_{0}$ with $\Phi_{0}=h/e$), the additional hopping phase arising from the external magnetic field is $\phi_{\mathbf{i},\mathbf{i}+\hat{\mathbf{x}}}=\Phi/\Phi
_{0}$ and $\phi_{\mathbf{i},\mathbf{i}+\hat{\mathbf{y}}}=0$. Here, $E_{F}$ is the Fermi energy (pinned by the superconductor condensate), $a$ is the lattice constant, and $A$, $B$, $C$, $D$, and $M$ are system parameters which can be experimentally controlled. The Hamiltonian $H_{\scriptscriptstyle{\mathbf{S}}}$ of the two-dimensional superconducting leads is: $H_{\scriptscriptstyle{\mathbf{S}}}=\sum_{\scriptscriptstyle{\mathbf{S}}\mathbf{{k}},\sigma}\epsilon
_{\scriptscriptstyle{\mathbf{S}}\mathbf{{k}}}a_{{\scriptscriptstyle{\mathbf{S}}}\mathbf{k}\sigma}^{\dagger} a_{{\scriptscriptstyle{\mathbf{S}}}\mathbf{k}\sigma}+\sum_{\scriptscriptstyle{\mathbf{S}}\mathbf{k}}
\Delta(e^{i\theta_{\scriptscriptstyle{\mathbf{S}}}}
a_{{\scriptscriptstyle{\mathbf{S}}}\mathbf{k}\uparrow}^{\dagger}%
a_{{\scriptscriptstyle{\mathbf{S}}}-\mathbf{k}\downarrow}^{\dagger}+H.c.)$ where the sum over $\scriptstyle{\mathbf{S}}$ refers to the left and right superconducting leads and $a_{{\scriptscriptstyle{\mathbf{S}}}\mathbf{k}\sigma}^{\dagger}$ ($a_{{\scriptscriptstyle{\mathbf{S}}}\mathbf{k}\sigma}$) is the creation (annihilation) operator in the superconducting leads for electrons with the momentum $\mathbf{k}=(k_{x},k_{y})$. Here we consider a general $s$-wave
superconductor; $\Delta$ is the superconductor gap, and $\theta_{{\scriptscriptstyle{\mathbf{S}}}}$ represents the superconducting phase of the left or right superconducting lead.

Noted that, it is straightforward to compute all momentum-space Green's functions, as well as the self-energy functions of the superconducting leads, using the superconducting Hamiltonian given above. To simplify use of the
Landauer-B\"{u}tiker formula in real space, it necessary to transform all functions into the form appropriate to the tight-binding approximation. Following the same method used in previous work,\cite{Sun2} we show the result in Appendix A. Then, the coupling Hamiltonian $H_{T}$ becomes: $H_{T} =\sum_{\scriptscriptstyle{\mathbf{S}}\mathbf{i}_{y}}(a^{\dagger}_{{\scriptscriptstyle{\mathbf{S}}} \mathbf{i}_{y} \uparrow}, a_{{\scriptscriptstyle{\mathbf{S}}} \mathbf{i}_{y} \downarrow}) \mathbf{t}_{{\scriptscriptstyle{\mathbf{S}C}}} \Psi
_{\mathbf{i}_{y}} +H.c. $, where the operator $a_{{\scriptscriptstyle{\mathbf{S}}} \mathbf{i}_{y}\sigma} =\sum_{\mathbf{k}} e^{i k_{y} \mathbf{i}_{y} a} a_{{\scriptscriptstyle{\mathbf{S}}}%
\mathbf{k}\sigma}$ and $\mathbf{t}_{{\scriptscriptstyle{\mathbf{S}C}}} =
\left(
\begin{array}
[c]{llll}%
t_{{\scriptscriptstyle{\mathbf{S}}}s} & t_{{\scriptscriptstyle{\mathbf{S}}}p}
& 0 & 0\\
0 & 0 & -t_{{\scriptscriptstyle{\mathbf{S}}}s} &
-t_{{\scriptscriptstyle{\mathbf{S}}}p}%
\end{array}
\right) .$ Here the parameters $t_{{\scriptscriptstyle{\mathbf{S}}}s}$ and $t_{{\scriptscriptstyle{\mathbf{S}}}p}$ are the coupling strengths between the superconductors and the HgTe/CdTe QWs, which depends on the interface contact
potential and, experimentally, the quality of the coupling.

\subsection{Current formula through a superconducting lead by non-equilibrium Green's function}
The supercurrent through superconducting leads is carried by Cooper pairs. However, one does not have to introduce the creation operator for the Cooper pair in order to calculate the current through the superconducting lead. Generally, the supercurrent from the left or right superconducting lead to the central device can be calculated from the evolution of the number operator of the electrons in the superconducting lead:
\begin{eqnarray}\label{Current1}
 &&I_{\scriptscriptstyle{\mathbf{S}}} = -e\langle\frac{d N_{\scriptscriptstyle{\mathbf{S}}}}{dt}\rangle
 =\frac{ie}{\hbar}\bigg\langle\big[\sum_{\mathbf{i},\sigma}a^\dagger_{{\scriptscriptstyle{\mathbf{S}}}\mathbf{i}\sigma} a_{{\scriptscriptstyle{\mathbf{S}}}\mathbf{i}\sigma},H\big]\bigg\rangle\nonumber\\
 &&=\frac{ie}{\hbar}\sum_{\alpha,\mathbf{i}_y,\sigma} [t_{{\scriptscriptstyle{\mathbf{S}}}\alpha}\langle a^{\dagger}_{{\scriptscriptstyle{\mathbf{S}}} \mathbf{i}_y \sigma}c_{\alpha\mathbf{i}_y\sigma}\rangle
 - t^*_{{\scriptscriptstyle{\mathbf{S}}}\alpha}\langle c^{\dagger}_{\alpha\mathbf{i}_y\sigma}a_{{\scriptscriptstyle{\mathbf{S}}} \mathbf{i}_y \sigma}\rangle]\nonumber\\
 &&= \frac{e}{\hbar}\sum_{\alpha,\mathbf{i}_y} Tr[t_{{\scriptscriptstyle{\mathbf{S}}}\alpha}G^{<}_{\alpha {\scriptscriptstyle{\mathbf{S}}}}(t,t;\mathbf{i}_y,\mathbf{i}_y)-t^*_{{\scriptscriptstyle{\mathbf{S}}}\alpha}G^{<}_{ {\scriptscriptstyle{\mathbf{S}}}\alpha }(t,t;\mathbf{i}_y,\mathbf{i}_y)]\nonumber\\
 &&= \frac{e}{\hbar} Tr[\mathbf{\Gamma_z}\mathbf{G}^{<}_{{\scriptscriptstyle{C\mathbf{S}}}}(t,t){\bf t}_{{\scriptscriptstyle{\mathbf{S}C}}}-\mathbf{\Gamma_z} {\bf t}^\dagger_{{\scriptscriptstyle{\mathbf{S}C}}} \mathbf{G}^{<}_{ {\scriptscriptstyle{\mathbf{S}C}}}(t,t)],
\end{eqnarray}
where particle number operator $N_{{\scriptscriptstyle{\mathbf{S}}}}=\sum_{\mathbf{i}\sigma} a^{\dagger}_{{\scriptscriptstyle{\mathbf{S}}}\mathbf{i}\sigma}a_{{\scriptscriptstyle{\mathbf{S}}}\mathbf{i}\sigma}$ and $\mathbf{\Gamma_z}=\sigma_z\bigotimes I_{2\times2}$.

In order to obtain the Green's function in the above formula, the contour-ordering Green's functions $\mathbf{G}(\tau,\tau')$ is defined in the Nambu representation:
\begin{eqnarray}
&&\mathbf{G}_{{\scriptscriptstyle{C\mathbf{S}}}}(\tau,\tau';\mathbf{i}_y,\mathbf{i}'_y)=\nonumber\\
&&-i\left[ \begin{array}{ll}
\big\langle \mathcal{T} \Psi_{\mathbf{i}_y\uparrow}(\tau)a^{\dagger}_{{\scriptscriptstyle{\mathbf{S}}} \mathbf{i}'_y \uparrow}(\tau')\big\rangle
& \big\langle \mathcal{T} \Psi_{\mathbf{i}_y\uparrow}(\tau)a_{{\scriptscriptstyle{\mathbf{S}}} \mathbf{i}'_y \downarrow}(\tau')\big\rangle \\
\big\langle \mathcal{T} \Psi^{\dagger}_{\mathbf{i}_y\downarrow}(\tau)a^{\dagger}_{{\scriptscriptstyle{\mathbf{S}}} \mathbf{i}'_y \uparrow}(\tau')\big\rangle
& \big\langle \mathcal{T} \Psi^{\dagger}_{\mathbf{i}_y\downarrow}(\tau)a_{{\scriptscriptstyle{\mathbf{S}}} \mathbf{i}'_y \downarrow}(\tau')\big\rangle \end{array}\right],\\
&&\mathbf{G}_{{\scriptscriptstyle{\mathbf{S}C}}}(\tau,\tau';\mathbf{i}_y,\mathbf{i}'_y)=\nonumber\\
&&-i\left[ \begin{array}{ll}
\big\langle \mathcal{T} a_{{\scriptscriptstyle{\mathbf{S}}} \mathbf{i}_y \uparrow}(\tau)\Psi^{\dagger}_{\mathbf{i}'_y\uparrow}(\tau')\big\rangle
& \big\langle \mathcal{T} a_{{\scriptscriptstyle{\mathbf{S}}} \mathbf{i}_y \downarrow}(\tau)\Psi_{\mathbf{i}'_y\uparrow}(\tau')\big\rangle \\
\big\langle \mathcal{T} a^{\dagger}_{{\scriptscriptstyle{\mathbf{S}}} \mathbf{i}_y \uparrow}(\tau)\Psi^{\dagger}_{\mathbf{i}'_y\downarrow}(\tau')\big\rangle
& \big\langle \mathcal{T} a^{\dagger}_{{\scriptscriptstyle{\mathbf{S}}} \mathbf{i}_y \downarrow}(\tau)\Psi_{\mathbf{i}'_y\downarrow}(\tau')\big\rangle \end{array}\right],
\end{eqnarray}
where the contour-ordering operator, $\mathcal{T}$, orders all operators following it along the contour time $\tau$ loop, and $\Psi_{\mathbf{i}\sigma}=(c_{s\mathbf{i}\sigma}, c_{p\mathbf{i}\sigma})^T$. In this case, a general relation for the contour-ordered Green's function $\mathbf{G}_{ {\scriptscriptstyle{C\mathbf{S}}}} (\tau,\tau';\mathbf{i}_y,\mathbf{i}'_y)$ can be derived rather easily, either using the equation of motion technique or with a direct expansion of the scattering matrix, as the follows:
\begin{eqnarray}
&&\mathbf{G}_{ {\scriptscriptstyle{C\mathbf{S}}}}(\tau,\tau';\mathbf{i}_y,\mathbf{i}'_y)=\sum_{{\bf j}_{y}}\int d\tau_1 \mathbf{G}_{ {\scriptscriptstyle{CC}}}(\tau,\tau_1;\mathbf{i}_y,{\bf j}_y)\nonumber\\
&&\times \mathbf{t}^\dagger_{\scriptscriptstyle{\mathbf{S}C}} \mathbf{g}_{{\scriptscriptstyle{\mathbf{SS}}}}(\tau_1,\tau';{\bf j}_y,\mathbf{i}'_y).
\end{eqnarray}
Here $\mathbf{G}_{\scriptscriptstyle{CC}}(\tau,\tau';\mathbf{i}_y,\mathbf{i}'_y)$ and $\mathbf{g}_{{\scriptscriptstyle{\mathbf{SS}}}}(\tau,\tau';\mathbf{i}_y,\mathbf{i}'_y)$ represent the contour-ordered Green's functions of the central region and the decoupled superconducting lead in the Nambu representation, respectively. Using the standard analytic continuation formula\cite{APJauhoPRB1994l} $C^<(t, t') =\int dt_1[A^r(t, t_1)B^<(t_1, t') + A^<(t, t_1)B^a(t_1, t')]$, the lesser Green's function in the Nambu representation is found to be

\begin{eqnarray}\label{Glesser1}
&&\mathbf{G}^<_{\scriptscriptstyle{C\mathbf{S}}}(t,t';\mathbf{i}_y,\mathbf{i}'_y)=\sum_{{\bf j}_{y}}\int dt_1 \nonumber\\
&&\mathbf{G}^r_{\scriptscriptstyle{CC}}(t,t_1;\mathbf{i}_y,{\bf j}_y) \mathbf{t}^\dagger_{\scriptscriptstyle{\mathbf{S}C}} \mathbf{g}^<_{{\scriptscriptstyle{\mathbf{SS}}}}(t_1,t';{\bf j}_y,\mathbf{i}'_y)\nonumber\\
&&+\mathbf{G}^<_{\scriptscriptstyle{CC}}(t,t_1;\mathbf{i}_y,{\bf j}_y) \mathbf{t}^\dagger_{\scriptscriptstyle{\mathbf{S}C}} \mathbf{g}^a_{{\scriptscriptstyle{\mathbf{SS}}}}(t_1,t';{\bf j}_y,\mathbf{i}'_y).
\end{eqnarray}
Similarly, $\mathbf{G}^<_{\scriptscriptstyle{\mathbf{S}C}}$ can be also obtained.
Then, substituting the above expression for $\mathbf{G}^<$ into Eq. (\ref{Current1}), the current is
\begin{eqnarray}\label{Current2}
 &&I_{{\scriptscriptstyle{\mathbf{S}}}}
 = \frac{e}{\hbar}\int dt_1
 Tr[\mathbf{\Gamma_z}\mathbf{G}^r_{\scriptscriptstyle{CC}}(t,t_1)
 \mathbf{t}^\dagger_{\scriptscriptstyle{\mathbf{S}C}} \mathbf{g}^<_{{\scriptscriptstyle{\mathbf{SS}}}}(t_1,t)\mathbf{t}_{{\scriptscriptstyle{\mathbf{S}C}}}\nonumber\\
 &\ &+\mathbf{\Gamma_z}\mathbf{G}^<_{\scriptscriptstyle{CC}}(t,t_1) \mathbf{t}^\dagger_{\scriptscriptstyle{\mathbf{S}C}} \mathbf{g}^a_{{\scriptscriptstyle{\mathbf{SS}}}}(t_1,t)\mathbf{t}_{{\scriptscriptstyle{\mathbf{S}C}}}]
 +H.c.\nonumber\\
 &&=\frac{e}{\hbar}\int dt_1 Tr[\mathbf{\Gamma_z}\mathbf{G}^r_{\scriptscriptstyle{CC}}(t,t_1)
 \mathbf{\Sigma}^<_{{\scriptscriptstyle{\mathbf{SS}}}}(t_1,t)\nonumber\\
 &\ &+\mathbf{\Gamma_z}\mathbf{G}^<_{\scriptscriptstyle{CC}}(t,t_1) \mathbf{\Sigma}^a_{{\scriptscriptstyle{\mathbf{SS}}}}(t_1,t)]
 +H.c.,
\end{eqnarray}
where $\mathbf{\Sigma}_{{\scriptscriptstyle{\mathbf{SS}}}}=\mathbf{t}^\dagger_{\scriptscriptstyle{\mathbf{S}C}} \mathbf{g}_{{\scriptscriptstyle{\mathbf{SS}}}}\mathbf{t}_{{\scriptscriptstyle{\mathbf{S}C}}}$ represents the self-energy function of the supercondcuting lead. In the case of zero bias voltage, because both the Green's function $\mathbf{G}_{\scriptscriptstyle{CC}}(t,t')$ and the self-energy function $\mathbf{\Sigma}_{{\scriptscriptstyle{\mathbf{SS}}}}(t,t')$ become functions of $t-t'$,\cite{QFSunPRB1999p} Eq. (\ref{Current2}) can be simplified as:
\begin{eqnarray}\label{Current3}
I_{{\scriptscriptstyle{\mathbf{S}}}}
&=&\frac{e}{h}\int dE Tr[\mathbf{\Gamma_z}\mathbf{G}^r_{\scriptscriptstyle{CC}}(E)
 \mathbf{\Sigma}^<_{{\scriptscriptstyle{\mathbf{SS}}}}(E)
 +\mathbf{\Gamma_z}\mathbf{G}^<_{\scriptscriptstyle{CC}}(E)\mathbf{\Sigma}^a_{{\scriptscriptstyle{\mathbf{SS}}}}(E)\nonumber\\
 &\ &-\mathbf{\Gamma_z}
 \mathbf{\Sigma}^<_{{\scriptscriptstyle{\mathbf{SS}}}}(E)\mathbf{G}^a_{\scriptscriptstyle{CC}}(E)
 -\mathbf{\Gamma_z}\mathbf{\Sigma}^r_{{\scriptscriptstyle{\mathbf{SS}}}}(E)\mathbf{G}^<_{\scriptscriptstyle{CC}}(E)].
\end{eqnarray}

The self-energy functions $\mathbf{\Sigma}^r_{{\scriptscriptstyle{\mathbf{SS}}}}$, $\mathbf{\Sigma}^a_{{\scriptscriptstyle{\mathbf{SS}}}}$ and $\mathbf{\Sigma}^<_{{\scriptscriptstyle{\mathbf{SS}}}}$ for the superconducting leads can be calculated as in Refs.[\onlinecite{Sun2,Sun3,Sun4,MPLSancho,QFSunPRB1999p}]. For the sake of completeness, surface Green's functions and retarded or lesser self-energy function of superconducting leads are given in Appendix A.

\begin{figure}[ptb]
\includegraphics[width=1.0\columnwidth]{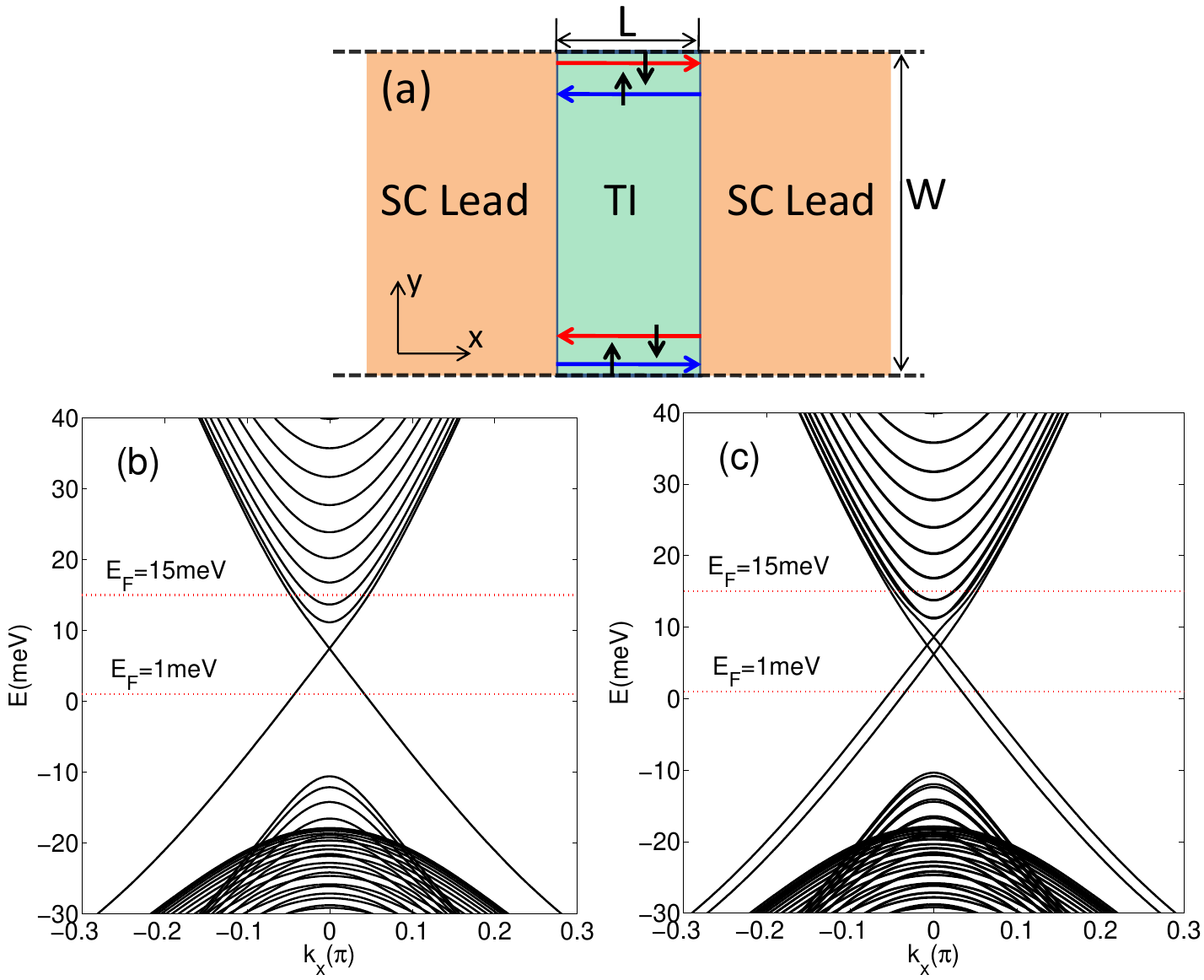}
\caption{(Color online) (a) Schematic diagram for the superconductor-TI-superconductor device. (b) and (c) show the energy spectra of the ribbon of HgTe/CdTe quantum wells in the absence of a magnetic field and in the presence of a magnetic field with the hopping phase $\Phi=0.001$. The ribbon width is $W=400nm$ and the effective
mass is $M=-10meV$.}
\label{Figure1}
\end{figure}

The Green's function of the central region can be calculated from
\begin{eqnarray}\label{TIGreenfunction}
{\bf G}^{r/a}_{\scriptscriptstyle{CC}}(E) =\{EI-{\bf H}_C -{\bf\Sigma}^{r/a}_{{\scriptscriptstyle{\mathbf{L}}}} -{\bf\Sigma}^{r/a}_{{\scriptscriptstyle{\mathbf{R}}}} \pm i\gamma I\}^{-1},
\end{eqnarray}
where $\mathbf{H}_{C}$ is the Hamiltonian of the central TI region as shown in Fig.\ref{Figure1}(a) (light blue region) and $\gamma$ represents the linewidth function of states, which is always chosen to be an infinitesimal quantity for clean samples. Consequently, the current through the left or right superconducting lead can be obtained straightforwardly using Eq. (\ref{Current3}). In the following numerical calculations, we choose parameters appropriate for the realistic materials:\cite{KonigScience2007r} (1) the HgTe/CdTe QW parameters are $A=364.5meVnm$, $B=-686meVnm^{2}$, $C=0$, and $D=-512meVnm^{2}$; (2) the superconductor parameters are the gap energy $\Delta=1meV$, the superconducting phases $\theta_{L}=\pi/2$ and $\theta_{R}=0$. The lattice constant $a$ is set to $5nm$ and the TI-S coupling strengths are taken to be $t_{{\scriptscriptstyle{\mathbf{S}}}s}= t_{{\scriptscriptstyle{\mathbf{S}}}p}\equiv t=3.5meV$.

Besides, it is worth pointing out that actually we do not need the full Green's function of the whole central region, and a recursive algorithm about how to obtain parts of retarded Green's function, used in Eqs. (\ref{Current3}), (\ref{LCurrent5}) and (\ref{DistributionofCooperpair1}), is briefly introduced in Appendix B.

\subsection{Local current formula and local Cooper pair distribution by non-equilibrium Green's function}
In order to check the real-space distribution of the supercurrent, it is necessary to derive a local current formula perpendicular to the cross section of the HgTe/CdTe QWs. Using the non-equilibrium Green's function method, the
current flowing through site $\mathbf{i}$ of the central region can be written as\cite{APJauhoPRB1994l}

\begin{eqnarray}\label{LCurrent1}
 &&J_{\mathbf{i}\uparrow} (t)= -e\langle\frac{d N_{\mathbf{i}\uparrow}}{dt}\rangle
 =\frac{ie}{\hbar}\bigg\langle\big[\sum_{\alpha\in s, p}c^\dagger_{\alpha\bf i\uparrow} c_{\alpha\bf i\uparrow},H\big]\bigg\rangle\nonumber\\
 &&=\frac{ie}{\hbar}\sum_{\alpha,\beta,\hat{\mathbf{r}}} [h_{\mathbf{i}\bf i+\hat{\mathbf{r}},\alpha\beta }\langle c^{\dagger}_{\alpha \mathbf{i} \uparrow}c_{\beta{\bf i}+\hat{\mathbf{r}}\uparrow}\rangle
 - h^*_{\mathbf{i}\mathbf{i}+\hat{\mathbf{r}},\alpha\beta }\langle c^\dagger_{\beta{\bf i}+\hat{\mathbf{r}}\uparrow}c_{\alpha \mathbf{i} \uparrow}\rangle]\nonumber\\
 &&=\frac{e}{\hbar}\sum_{\hat{\mathbf{r}}} Tr[h_{\mathbf{i}\mathbf{i}+\hat{\mathbf{r}}}\mathbf{G}^{<}_{\scriptscriptstyle{CC},\uparrow\uparrow}(t,t;{\bf i+\hat{\mathbf{r}}},{\bf i})
 +H.c.]\nonumber\\
 &&= \frac{e}{\hbar}\sum_{\hat{\mathbf{r}}} Tr[H_{\mathbf{i}\mathbf{i}+\hat{\mathbf{r}}}\mathbf{G}^{<}_{\scriptscriptstyle{CC}}(t,t;{\bf i+\hat{\mathbf{r}}},{\bf i})
 +H.c.]_{\uparrow\uparrow}\nonumber\\
 &&= \frac{2e}{\hbar} Re\sum_{\hat{\mathbf{r}}} Tr[H_{\mathbf{i}\mathbf{i}+\hat{\mathbf{r}}}\mathbf{G}^{<}_{\scriptscriptstyle{CC}}(t,t;{\bf i+\hat{\mathbf{r}}},{\bf i})]_{\uparrow\uparrow}
\end{eqnarray}
and similarly in the Nambu representation,
\begin{eqnarray}\label{LCurrent2}
J_{\mathbf{i}\downarrow}(t) = -\frac{2e}{\hbar} Re\sum_{\hat{\mathbf{r}}} Tr[H_{\mathbf{i}\mathbf{i}+\hat{\mathbf{r}}}\mathbf{G}^{<}_{\scriptscriptstyle{CC}}(t,t;{\bf i+\hat{\mathbf{r}}},{\bf i})]_{\downarrow\downarrow}.
\end{eqnarray}
Here, the Hamiltonian matrices are the same to those defined in the previous section and the Keldysh lesser Green's function of the central region is defined as:
\begin{eqnarray}\label{LCurrent3}
&&\mathbf{G}^<_{\scriptscriptstyle{CC}}(t,t';\mathbf{i},\mathbf{j}) = i\langle\Psi^{\dagger}_\mathbf{j}(t')\Psi_\mathbf{i}(t)\rangle.
\end{eqnarray}

If we set the voltage of the superconducting leads to zero ($V_{L}=V_{R}=0$), it can be shown that all Green's functions of the central region depend only on the time difference; namely $\mathbf{G}_{\scriptscriptstyle{CC}} (t,t^{\prime};\mathbf{i},\mathbf{j})=\mathbf{G}_{\scriptscriptstyle{CC}} (t-t^{\prime},0;\mathbf{i},\mathbf{j})$.\cite{QFSunPRB1999p} After taking the Fourier transform of the lesser Green's function, the local current between neighboring sites $\mathbf{i}$ and $\mathbf{j}=\mathbf{i}+\hat{\mathbf{r}}$ can be calculated from the formula
\begin{eqnarray}\label{LCurrent4}
J_{\mathbf{i}\rightarrow \mathbf{j}}&=&\frac{2e^2}{\hbar}Re\big\{Tr[\mathbf{\Gamma_z}H_{\mathbf{i}\mathbf{j}} \mathbf{G}_{\scriptscriptstyle{CC}}^{<}(0,0;\mathbf{j},\mathbf{i})]\big\}\nonumber\\
&=&\frac{2e^2}{h}Re\int^{\infty}_{-\infty}dE\ Tr[\mathbf{\Gamma}_zH_{\mathbf{i}\mathbf{j}}\mathbf{G}_{\scriptscriptstyle{CC}}^{<}(E;\mathbf{j},\mathbf{i})].
\end{eqnarray}
By applying the Keldysh equation $\mathbf{G}^<=\mathbf{G}^r(i\mathbf{\Gamma}_Lf_L+i\mathbf{\Gamma}_Rf_R)\mathbf{G}^a$,\cite{APJauhoPRB1994l} Eq. (\ref{LCurrent4}) can be simplified to
\begin{eqnarray}\label{LCurrent5}
J_{\mathbf{i}\rightarrow \mathbf{j}}&=&-\frac{2e^2}{h}Im\int^{0}_{-\infty}dE \nonumber\\
&\ & \times Tr\big\{\mathbf{\Gamma}_zH_{\mathbf{i}\mathbf{j}}
[\mathbf{G}_{\scriptscriptstyle{CC}}^r(\mathbf{\Gamma}_L+\mathbf{\Gamma}_R)\mathbf{G}_{\scriptscriptstyle{CC}}^a]_{\mathbf{j}\mathbf{i}}\big\},
\end{eqnarray}
where the linewidth functions of superconducting leads are $\mathbf{\Gamma}_{L/R}= i(\mathbf{\Sigma}^r_{L/R}-\mathbf{\Sigma}^a_{L/R})$, and the Fermi energies of the left and right superconducting leads are set to zero, and thus $f_L=f_R=1/0$ below/above the Fermi energy. Thus, given that the retarded and advanced Green's functions are known, the local current can be readily calculated using the above formula.

In addition, the local Cooper pair distribution can be approximately obtained by the computing superconductor order parameter,
\begin{eqnarray}\label{DistributionofCooperpair1}
&&\langle c^\dagger_{\mathbf{i}\uparrow}c^\dagger_{\mathbf{i}\downarrow}\rangle=Im\int^{\infty}_{-\infty}\frac{dE}{2\pi}
Tr\big\{[\mathbf{G}_{\scriptscriptstyle{CC}}^<(E,\mathbf{i},\mathbf{i})]_{\uparrow,\downarrow}\big\} \nonumber\\
&&=Re\int^{\infty}_{-\infty}\frac{dE}{2\pi} Tr\big\{[\mathbf{G}_{\scriptscriptstyle{CC}}^r(\mathbf{\Gamma}_Lf_L+\mathbf{\Gamma}_Rf_R)
\mathbf{G}_{\scriptscriptstyle{CC}}^a]_{\mathbf{i}\uparrow,\mathbf{i}\downarrow}\big\},\nonumber\\
&&=Re\int^{0}_{-\infty}\frac{dE}{2\pi} Tr\big\{[\mathbf{G}_{\scriptscriptstyle{CC}}^r (\mathbf{\Gamma}_L+\mathbf{\Gamma}_R) \mathbf{G}_{\scriptscriptstyle{CC}}^a]_{\mathbf{i}\uparrow,\mathbf{i}\downarrow}\big\},
\end{eqnarray}
where $V_L=V_R=0$ has been adopted.

\section{Results and Discussions}~\label{sec:discussions}
~\label{sec:discussions}
In order to facilitate comparison of our results with previous experimental measurements, we have performed calculations for essentially the same device as used in the experiments. As shown in Fig. {\ref{Figure1}}, the Josephson junction consists of the central region of HgTe/CdTe QWs and two superconducting leads. Applying a perpendicular magnetic field, a supercurrent interference pattern is expected to be observed as the magnetic strength is varied. In order to discuss the following calculation results explicitly, the energy dispersion in the absence and presence of external magnetic field are given in Fig. 1(b) and 1(c), respectively. Note that the ribbon width is set to $W=400nm$, and the effective mass to $M =-10meV$ which corresponds to the HgTe/CdTe QW being a TI. In addition, the
hopping phase arising from the external magnetic field in Fig. 1(c) is set to $\Phi=0.001$.

One of the most intriguing observations reported in one recent experiment\cite{HartNautrePhys2014f} is that although there is still residual resistance from bulk states, it seems that only topologically protected edge states contribute to the supercurrent interference pattern while the bulk states appear to be `frozen' in the transport measurement. One of the underlying reasons may be the special characteristics of TIs, which result in the edge states predominating in the proximity effect to a superconductor, while at the same time suppressing the contribution from the bulk states. If
this supposition is true, TI-Superconductor hybrid devices could open a fascinating new chapter in this field; if not, a reasonable explanation of the experimental results is still necessary.

\begin{figure}[ptb]
\includegraphics[width=1.0\columnwidth]{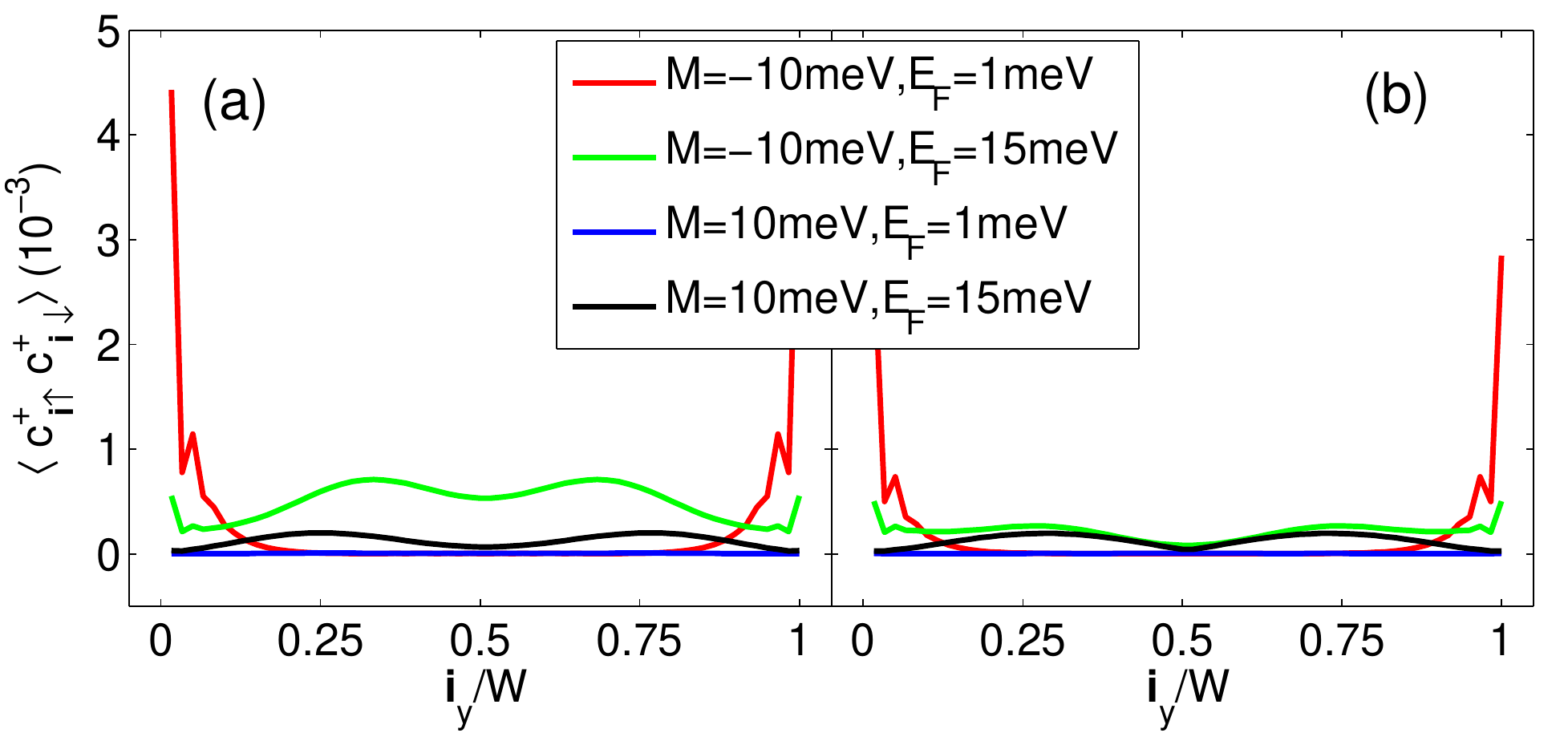}
\caption{(Color online) (a) and (b) are the superconductor order parameters in the central TI region in the absence of a magnetic field and in the presence of a magnetic field ($\Phi=0.001$), respectively. Noted that $\mathbf{i}_{y}$ represents the $y$ index of the site located at the middle cross section of HgTe/CdTe QWs. The ribbon width is $W=300nm$ and the length of the central region is $L=400nm$. }\label{Figure2}
\end{figure}

An effective and feasible approach to verify this assumption, that the system topology has endowed topologically protected edge states with some special characteristics but does not do the same for bulk states, is to examine the
real-space distribution of Cooper pairs along the cross section through the central TI sample. When $M=-10meV$ and $E_{F}=1meV$ as shown in Fig. \ref{Figure2}(a), Cooper pairs localize mainly near the two edges of the Josephson junction. This distribution origins with the topologically protected edge states being confined on the sample edges. As the Fermi energy is shifted to the conduction band edge, $E_{F}=15meV$, Cooper pairs spread over the whole cross section of the QWs, which indicates that the bulk states become dominant. When QWs are tuned to be non-topological, $M=10meV$, the occupation number of Cooper pairs decreases to zero for $E_{F}=1meV$ because there is no state within the band gap. Similarly, Cooper pairs spread over all the whole cross section of QWs when the Fermi energy is again tuned to the edge of conductance band, $E_{F}=15meV$ in Fig. \ref{Figure2}(a). There is no change in essence when an external magnetic field is applied, as shown in Fig. \ref{Figure2}(b).

From Fig. \ref{Figure2}, it can be concluded that there is no essential difference between topologically-protected edge states and bulk states in proximity to a superconductor. Even though topologically-protected edge states, because of their high mobility, could have some advantages in proximity to a superconductor, the bulk states are not forbidden to contribute and can experience the proximity effect. This conclusion is still correct when an external magnetic field is turned on. This does not resolve the problem of why, in the experiment\cite{HartNautrePhys2014f} mentioned above, only topologically-protected edge states appear to contribute to the supercurrent interference pattern even though there are residual bulk states present.

\begin{figure}[ptb]
\includegraphics[width=0.8\columnwidth]{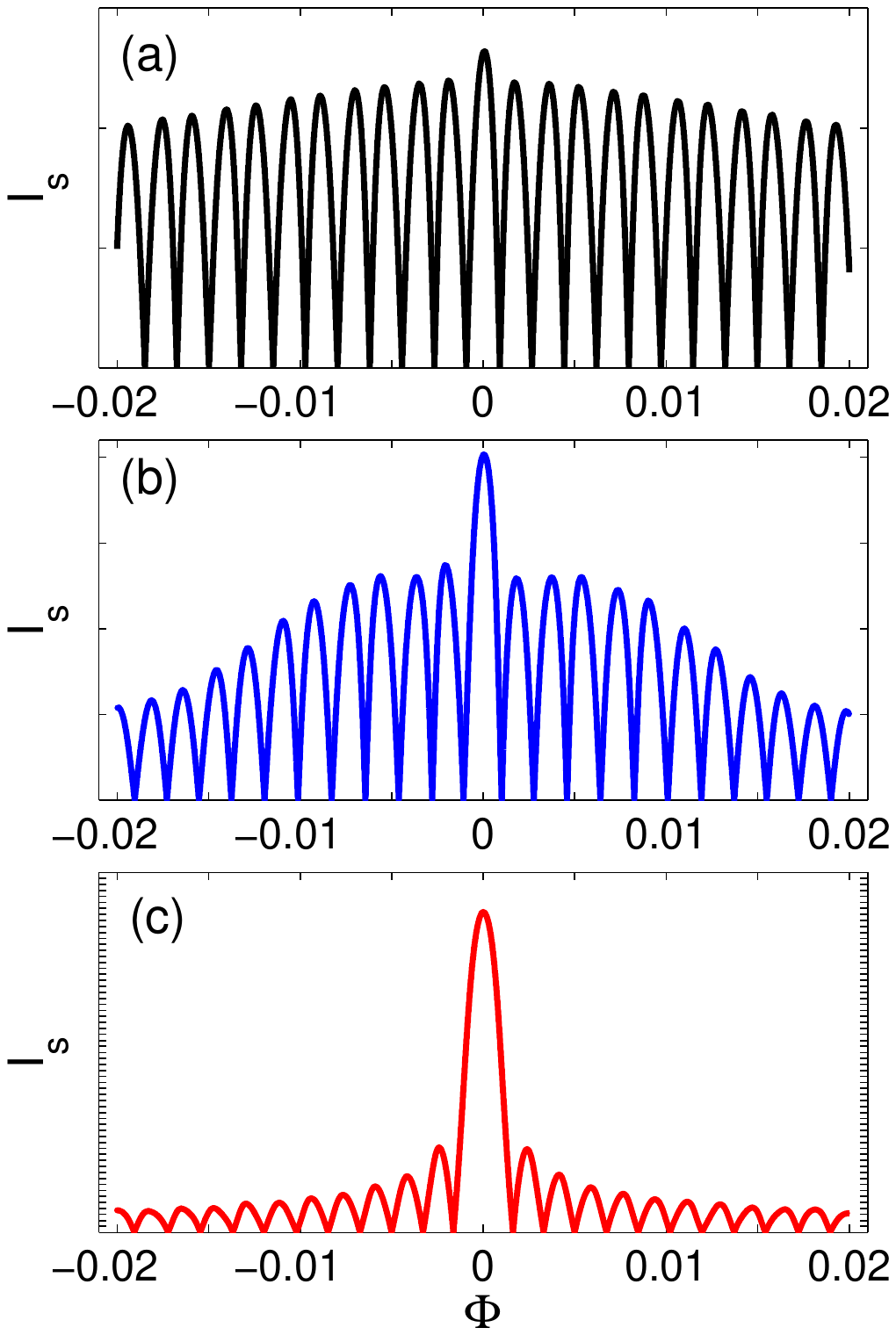}
\caption{(Color online)Interference pattern shown in critical current for the HgTe/CdTe hybrid system. (a), (b) and (c) are for the Fermi energies of $E_{F}=1meV$, $E_{F}=12meV$ and $E_{F}=60meV$, respectively. The ribbon width is $W=500nm$, the length of the central region is $L=100nm$, and the effective mass is $M=-10meV$. Note that $\Phi$ represents the additional hopping phase due to the external magnetic field, which is defined in Sec. \ref{sec:Hamiltonians}.} \label{Figure3}
\end{figure}

To understand the connection between our results and the experimental results,\cite{HartNautrePhys2014f} we present interference patterns for HgTe/CdTe hybrid system from the perspective of numerical simulation. This will illuminate the underlying physics. When the Fermi energy is set to $E_{F}=1meV$ in Fig. \ref{Figure3}(a), which means there are only two pairs of topologically-protected edges states confined to the top and bottom boundaries of the central sample, an interference pattern similar to a two-slit interference pattern is observed. In particular, when counting period of the fluctuations in Fig.\ref{Figure3}(a), it is found that they do correspond to a change of $\Phi_{0}/2$ in the magnetic flux threading the whole area of the central HgTe/CdTe QWs. When the Fermi energy is shifted up to $E_{F}=12meV$, the perfect two-slit interference pattern is damaged because of the addition of some bulk states, as shown in Fig.\ref{Figure3}(b). Note first that the heights of the interference peaks decrease gradually as the threading magnetic flux increases. Secondly, the modulation period of the interference pattern is not exactly equal to the change of the magnetic flux $\Phi_{0}/2$ threading the whole area of the central HgTe/CdTe QWs, and in particular the width of the central lobe clearly increases in size. As the Fermi energy is further moved up to $E_{F}=60meV$, an interference pattern similar to a single-slit Frauhofer interference pattern is obtained, as seen in Fig. 3(c). In this case, many bulk states mask the role of the two pairs of edge states in the interference. Besides, it was reported in Ref. [\onlinecite{GTkachov2015k}] that both the temperature and the length of Josephson junction can influence the shape of interference pattern obviously.

Comparing our theoretical results to the experimental ones reported in Ref. [\onlinecite{HartNautrePhys2014f}], it is easy to see that there are always residual bulk states even for the perfect `double-slit interference pattern', namely Fig. 2(c) in Ref. [\onlinecite{HartNautrePhys2014f}], while the authors thought there were only edge states. The question mentioned in the introduction and elsewhere therefore arises as to why no bulk states and only edge states are found when transforming the interference pattern into a real-space current distribution along the cross section of the HeTe/CdTe QWs using the Dynes-Fulton method.\cite{RCDynesPRB1971d}

\begin{figure}[ptb]
\includegraphics[width=0.8\columnwidth]{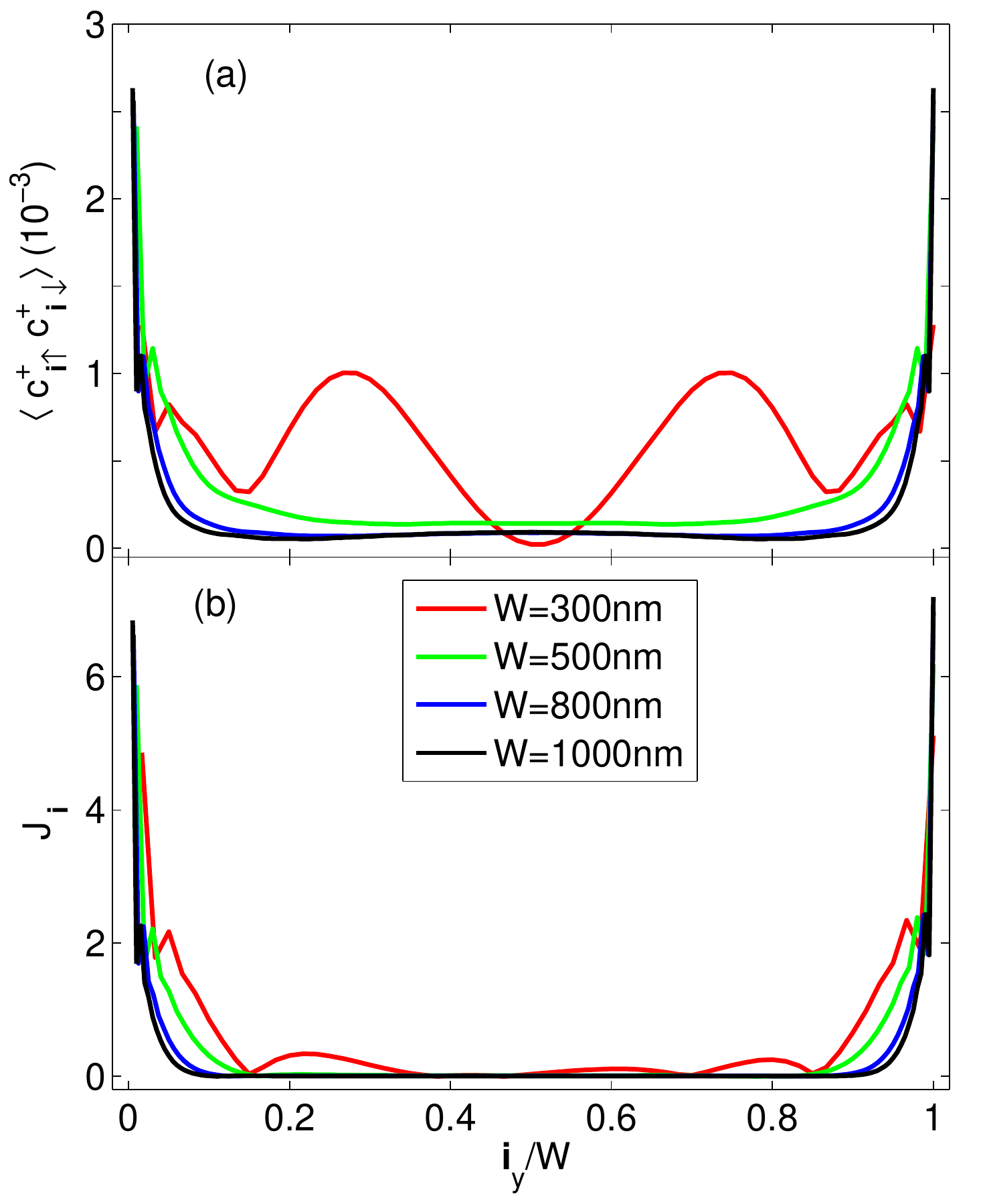}
\caption{(Color online) (a) Distribution of Cooper pair and (b) magnitude of local current along the cross section of the central TI sample, where $J_{\mathbf{i}}=|J_{\mathbf{i} \rightarrow\mathbf{i+\hat{x}}}|$, and $\mathbf{i}_{y}$ represents the $y$ index of the site locating at the middle cross section of the HgTe/CdTe QWs. Noted that a definite resistance (about $h/8e^2$) are preserved by choosing a suitable Fermi energy no matter what the width of sample is. The length of the central region is taken as $L=100nm$,and the effective mass is $M=-10meV$.}\label{Figure4}%
\end{figure}

One possible reason, which has been discussed in Ref. [\onlinecite{{HuiPRB2014g}}], is that the lack of bulk states may arise from the highly simplifying assumptions used in the Dynes-Fulton analysis of Ref. [\onlinecite{HartNautrePhys2014f}]. However, we want to present an alternate explanation of the experimental results. In Fig. \ref{Figure4}, we plot the distribution of Cooper pairs and supercurrent along the cross section as a function of the sample width. In order to compare our theoretical results conveniently with those of Ref. [\onlinecite{HartNautrePhys2014f}], here we maintain a system with a definite resistance (about $h/8e^{2}$)\cite{Interpretation1} by adjusting the Fermi energy appropriately no matter what the width of sample is in Fig. \ref{Figure4}.

It can be clearly seen from Fig. \ref{Figure4} that both the distribution of Cooper pairs and the magnitude of the local supercurrent at the sample center decrease gradually as the sample width increases. When the sample width is
increased to $W=1\mu m$, the contribution of the bulk states to Cooper pair or supercurrent is so small as to be negligible. However, the contribution of the edge states undergoes almost no change as the sample width changes from
$W=0.3\mu m$ to $W=1\mu m$, because the topological edge states are located in a finite range with characteristic decay length $\xi\sim\frac{A}{M}$.\cite{BZhouPRL2008j,HJiangPRL2014p} Moreover, the interference pattern is determined by the square of the local supercurrent density. Therefore, one clear result is that only Cooper pairs or supercurrent confined near the top and bottom boundaries of the TI are detectable experimentally. In fact, this was just the case in Ref. [\onlinecite{HartNautrePhys2014f}], where the sample width was $W=4\mu m$ and the residual resistance from bulk states
was approximately $h/6e^{2}$. Because the wave functions of these residual bulks states spread over the whole cross-section of the sample and thus are much weaker than those of two pairs of edge states which are always confined at the sample boundaries, to detect the contribution of finite residual bulk states to the supercurrent becomes very difficult experimentally. That is why only induced superconductivity in the quantum spin Hall edge states is observed in Ref. [\onlinecite{HartNautrePhys2014f}]. Through changing the TI sample width and tuning the Fermi energy by gate voltage, the results in Fig. \ref{Figure4} can be easily checked for experiments.

Finally, it should be noted that in order to obtain interference patterns similar to those seen in the experiment,\cite{HartNautrePhys2014f} a relative large linewidth, $\gamma_{\mathbf{i}}=1meV$, had to be used for every lattice site of the central TI sample. It is found that the interference pattern deviates significantly from that reported in the experiment\cite{HartNautrePhys2014f} if the linewidth $\gamma$ is set to be small. The reason is twofold: (1) the finite width of the central sample adopted in our simulation leads to discrete energy levels, therefore a relative large linewidth $\gamma_{\mathbf{i}}$ is needed to smear discrete energy levels and makes them look like continuum energy spectrum; (2) the experimental samples are somewhat disordered, which leads to broadening of states, however this supposition needs to be further confirmed experimentally. Since residual bulk states and dephasing are two significant factors influencing the robustness of the edge states of TIs, production of a clean, high quality TI sample is a long sought goal of this field. At present, residual bulk states are still always observed, no matter whether in 2D TI samples or in 3D ones. How to tune the Fermi energy completely away from the conduction/valence band edges into the band gap is still a big challenge before us.

\textcolor{red}{In addition, it should be explicitly pointed out that unusual interference patterns at low temperature, reported in Ref. [\onlinecite{GTkachov2015k}], is not observed in this work. Note that a requirement to obtain the unusual interference pattern is to resolve individual Andreev bound states, namely the resulting thermal smearing $k T$ is smaller than level spacing $\hbar /L$, where $L$ represents length of Josephson junction. In this work, though we work at zero temperature $T=0$, a finite linewidth for each site is used to simulate disorder effects, and also simulate a continuous continuum energy spectrum (namely a short junction with very large width). This assumption could hinder observation of unusual interference patterns in our work.}

\section{Conclusions}
~\label{sec:conclusions}
We have studied the transport properties of the superconductor-HeTe/CdTe QWs-superconductor Josephson junction in the presence of an applied magnetic field. It was found that the proximity effect can induce superconductivity for both edge and bulk states. Compared to bulk states, topologically-protected edge states do not possess any obvious advantages arising from being in close proximity to a superconductor. The fact that experiments have only observed induced superconductivity only in quantum spin Hall edge states is just due to the effect of size: the magnitude of the wave functions from bulk states become very small when these residual bulk states spread over the relatively large cross section of the sample ($W=4\mu m$).

From the results of this work, one meaningful thing is that even though there is limited residual bulk resistance for HgTe/CdTe QWs, it does not affect the superconducting quantum interference pattern for S-TI-S Josephson junction. However, on the other hand the experimental results tell us that residual bulk states are still present even after many years attempts in experiment, and the Fermi energy for HeTe/CdTe QWs does not lie inside the expected bulk gap. In this sense, there is still a long way to go to realize a perfect quantum spin Hall system, which is a prerequisite for obtaining useful topological devices for technological purposes.

\section*{ACKNOWLEDGMENTS}
We gratefully acknowledge the financial support from NSF-China under Grant Nos. 11204065(J.T.S.), 11474085(J.T.S.), 11504008(H.W.L.), 11574245(J.L.), 11274364(Q.F.S.), 11574007(Q.F.S.), and 11504008(X.C.X.), and NBRP of China under Grand Nos. 2012CB921303 and 2015CB921102. J.T.S. is also supported by NSF-Hebei Province under Grant No. BJ2014038.
\\

\section*{APPENDIX A}\label{AppA}
\setcounter{equation}{0}
\renewcommand{\theequation}{A.\arabic{equation}}

In this appendix, we intend to obtain the surface Green's function $\mathbf{g}^r_{\scriptscriptstyle{S}}$ of the  two-dimensional superconducting lead and the self-energies $\mathbf{\Sigma}^r_{\scriptscriptstyle{S}}$ associated with  the superconducting lead in real space. Note that a detailed derivation can be found in a published paper.\cite{Sun2}

Using the Hamiltonian of the superconducting lead introduced in Sec. \ref{sec:Hamiltonians}, the bulk Green's function $\mathbf{g}^r_{\scriptscriptstyle{S}}$ of the superconductor can be obtained readily in momentum space:
\begin{eqnarray}\label{AA1}
&&\mathbf{g}^r_{\scriptscriptstyle{S}}(\mathbf{k};E) =\frac{1}{(E+i0^+)\mathbf{I}_{2\times 2}-\mathbf{H}_{\scriptscriptstyle{S}}(\mathbf{k})}\nonumber\\
&&=A_\mathbf{k}^{-1}\left(
\begin{array}{cc}
E_++\varepsilon_\mathbf{k} & \Delta \\
\Delta & E_+-\varepsilon_\mathbf{k}
\end{array}
\right),
\end{eqnarray}
where $A_\mathbf{k}=E_+^2-\varepsilon_\mathbf{k}^2-\Delta^2$ and $E_+=E+i 0^+$. Here, the space of the above matrix refers to the wavevectors $|k\uparrow\rangle$ and $|-k\downarrow\rangle$. Using the Fourier transformation, we can write the surface Green's function of  two-dimensional superconducting leads in real space\cite{Sun2}
\begin{eqnarray}\label{AA2}
\mathbf{g}^r_{\scriptscriptstyle{S}}(y,y';E) =-i\pi\rho J_0(k_F(y-y'))\beta(E)\left(
\begin{array}{cc}
1 & \Delta/E \\
\Delta/E & 1
\end{array}
\right).\nonumber\\
\end{eqnarray}
where $J_0(x)$ is the Bessel function of the first kind, $\beta(E)=|E|/\sqrt{E^2-\Delta^2}$ for $|E|>\Delta $ and $\beta(E)=E/(i\sqrt{\Delta^2 -E^2})$ for $|E| < \Delta$. In addition, the density of electron states $\rho(\varepsilon_k)=\rho$ is assumed to be independent of the superconducting spectrum $\varepsilon_k$, which makes the detailed expression for $\varepsilon_k$ unnecessary.\cite{Sun2}

Straightforwardly, using the surface Green's function $\mathbf{g}^r_{\scriptscriptstyle{S}}$ for the superconducting lead, the self-energies $\mathbf{\Sigma}_{\scriptscriptstyle{S}}$ take the form:
\begin{eqnarray}\label{AA3}
&&\mathbf{\Sigma}^r_{{\scriptscriptstyle{S}},nm}(E) =t_c\mathbf{g}^r_S(y_n,y_m;E)t^*_c \nonumber\\
&&\ \ \ \ \ \ \ =-i\pi|t_c|^2\rho J_0(k_F(y_n-y_m))\beta(E)\left(
\begin{array}{cc}
1 & \Delta/E \\
\Delta/E & 1
\end{array}\"{u}
\right)\nonumber\\
&&\ \ \ \ \ \ \ =-\frac{i}{2}\Gamma_{{\scriptscriptstyle{S}},nm}(E),\nonumber\\
&&\mathbf{\Sigma}^a_{{\scriptscriptstyle{S}},nm}(E)=\frac{i}{2}\Gamma^*_{{\scriptscriptstyle{S}},nm}(E),\nonumber\\
&&\mathbf{\Sigma}^<_{{\scriptscriptstyle{S}},nm}(E)=-f(E)(\mathbf{\Sigma}^r_{{\scriptscriptstyle{S}},nm}-\mathbf{\Sigma}^a_{{\scriptscriptstyle{S}},nm}),\\
&&\mathbf{\Sigma}^>_{{\scriptscriptstyle{S}},nm}(E)=[1-f(E)](\mathbf{\Sigma}^r_{{\scriptscriptstyle{S}},nm}-\mathbf{\Sigma}^a_{{\scriptscriptstyle{S}},nm}),\nonumber
\end{eqnarray}
where $t_c$ represent the coupling strength between the supperconducting leads and the central sample $t_{{\scriptscriptstyle{\mathbf{S}}}s}$ or $t_{{\scriptscriptstyle{\mathbf{S}}}p}$, and $f(E)$ is the Fermi function. If we suppose now that the couplings of the superconducting leads to the $s$ and $p$ orbitals in the HgTe/CdTe QWs are identical, the final self energy takes the form $\mathbf{\Sigma}_{\scriptscriptstyle{S}} \otimes \mathbf{I}_{2\times 2}$.

\section*{APPENDIX B}\label{AppB}
\setcounter{equation}{0}
\renewcommand{\theequation}{B.\arabic{equation}}
In this appendix, we present a numerical method about how to compute the retarded Green's function of the central region in Eq. (\ref{Current3}) or (\ref{LCurrent5}).

A straightforward method to obtain the retarded Green's function of the central region is via a direct inversion of large matrix, as shown in Eq. (\ref{TIGreenfunction}). However, the direct inversion of large matrix is greatly time-consuming for computers. In order to obtain a current by Landauer-B\"{u}ttiker formula, in fact we only need parts of the retarded Green's function of the central region, e.g. that of the left-most layer contacting with the left lead, that of the right-most layer contacting with the right lead, that between the left-most layer and the right-most layer, and that of one specific layer in the central region, which can be represented as $G^r_{11}$, $G^r_{NN}$, $G^r_{1N}$ or $G^r_{N1}$, and $G^r_{nn}$, respectively.

Using Dyson equation, it can be easily proved that hopping Green's function between the left-most layer and the right-most layer as well as that of the right-most layer can be obtained by the following recursive algorithm:
\begin{eqnarray}\label{BB1}
&&At\ beginning:\nonumber\\
&&\ \ \ \ G^r_{NN}=1/(E-H_{11}-\Sigma^r_L)\nonumber\\
&&\ \ \ \ G^r_{1N}=G^r_{NN}\nonumber\\
&&Iterative\ process:\nonumber\\
&&\ \ \ \ for\ i=2:1:N\nonumber\\
&&\ \ \ \ \ \ G^r_{NN}=1/(E-H_{11}-H_{21}G^r_{NN}H_{12})\ \ \ \ \ \ \ \ \ \ \ \ \nonumber\\
&&\ \ \ \ \ \ G^r_{1N}=G^r_{1N}H_{12}G^r_{NN}\nonumber\\
&&At\ ending:\nonumber\\
&&\ \ \ \ G^r_{NN}=1/(1/G^r_{NN}-\Sigma^r_{R})\nonumber\\
&&\ \ \ \ G^r_{1N}=G^r_{1N}+G^r_{1N}\Sigma^r_{R}G^r_{NN}\nonumber
\end{eqnarray}
where $H_{11}$, $H_{12}$, $H_{21}(=H^{\dagger}_{12})$ represents the Hamiltonian of an arbitrary layer, the hopping Hamiltonians to the right neighboring layer and the left neighboring layer, respectively. $\Sigma^r_L$ and $\Sigma^r_R$ are self energies from left lead and right lead. If there are many sites for one layer, all quantities above will become a matrix with the same dimension to the site number in one layers. It should be pointed out explicitly that $G^r_{1N}$ and $G^r_{NN}$ are, namely, Green's function from the left-most layer to the right-most one and that of the right-most layer contacting with the left lead, respectively. Similarly, we can obtain $G^r_{N1}$ and $G^r_{11}$ if $\Sigma^r_L$ and $\Sigma^r_R$ commute each other, as well as $H_{12}$ and $H_{21}$.

From the recursive algorithm above, the Green's functions of one specific layer in the central region can be obtained as follows:
\begin{eqnarray}\label{BB1}
&&At\ beginning:\nonumber\\
&&\ \ \ \ G^r_{ss}=1/(E-H_{11}-\Sigma^r_L)\nonumber\\
&&\ \ \ \ G^r_{1s}=G^r_{ss}\nonumber\\
&&\ \ \ \ G^r_{tt}=1/(E-H_{11}-\Sigma^r_R)\nonumber\\
&&\ \ \ \ G^r_{Nt}=G^r_{tt}\nonumber\\
&&Iterative\ process:\nonumber\\
&&\ \ \ \ for\ i=2:1:n-1\nonumber\\
&&\ \ \ \ \ \ G^r_{ss}=1/(E-H_{11}-H_{21}G^r_{ss}H_{12})\nonumber\\
&&\ \ \ \ \ \ G^r_{1s}=G^r_{1s}H_{12}G^r_{ss}\nonumber\\
&& \nonumber\\
&&\ \ \ \ for\ i=N-1:-1:n+1\nonumber\\
&&\ \ \ \ \ \ G^r_{tt}=1/(E-H_{11}-H_{12}G^r_{tt}H_{21})\nonumber\\
&&\ \ \ \ \ \ G^r_{Nt}=G^r_{Nt}H_{21}G^r_{tt}\nonumber\\
&&At\ ending:\nonumber\\
&&\ \ \ \ G^r_{nn}=1/(E-H_{11}-H_{21}G^r_{ss}H_{12}-H_{12}G^r_{tt}H_{21})\nonumber\\
&&\ \ \ \ G^r_{1n}=G^r_{1s}H_{12}G^r_{nn}\nonumber\\
&&\ \ \ \ G^r_{Nn}=G^r_{Nt}H_{21}G^r_{nn}.\nonumber
\end{eqnarray}
Here, $G^r_{nn}$, $G^r_{1n}$, and $G^r_{Nn}$ are the Green's function of the $n$-th layer, the hopping Green's function from the left-most layer to the $n$-th layer, and the hopping Green's function from the right-most layer to the $n$-th layer, respectively. Similarly, the hopping Green's functions from the $n$-th layer to the left-most layer and to the right-most layer can be also obtained through this recursive method.

\end{document}